\documentclass[aps,pra,amsmath,amssymb,reprint,superscriptaddress,tightenlines]{revtex4-2}

\usepackage{graphicx}
\usepackage{dcolumn}
\usepackage{bm}
\usepackage[utf8]{inputenc}
\usepackage[T1]{fontenc}
\usepackage{lmodern}
\usepackage{cleveref}
\usepackage{color}

\crefname{equation}{}{}

\begin{document}

\title{Quantum reservoir computation utilising scale-free networks}

\author{Akitada Sakurai}
\affiliation{School of Multidisciplinary Science, Department of Informatics, SOKENDAI (the Graduate University for Advanced Studies), 2-1-2 Hitotsubashi, Chiyoda-ku, Tokyo 101-8430, Japan}
\affiliation{National Institute of Informatics, 2-1-2 Hitotsubashi, Chiyoda-ku, Tokyo 101-8430, Japan}

\author{Marta P. Estarellas}
\altaffiliation[Current Affiliation: ]{Qilimanjaro Quantum Tech, Carrer dels Comtes de Bell-Lloc, 161, 08014 Barcelona, Spain}
\affiliation{National Institute of Informatics, 2-1-2 Hitotsubashi, Chiyoda-ku, Tokyo 101-8430, Japan}

 \author{William J. Munro}
 \affiliation{NTT Basic Research Laboratories \& Research Center for Theoretical Quantum Physics,  3-1 Morinosato-Wakamiya, Atsugi, Kanagawa, 243-0198, Japan}
 \affiliation{National Institute of Informatics, 2-1-2 Hitotsubashi, Chiyoda-ku, Tokyo 101-8430, Japan}
 
\author{Kae Nemoto}  \email{Corresponding author: nemoto@nii.ac.jp}
\affiliation{National Institute of Informatics, 2-1-2 Hitotsubashi, Chiyoda-ku, Tokyo 101-8430, Japan}
\affiliation{School of Multidisciplinary Science, Department of Informatics, SOKENDAI (the Graduate University for Advanced Studies), 2-1-2 Hitotsubashi, Chiyoda-ku, Tokyo 101-8430, Japan}

\begin{abstract}
Today’s quantum processors composed of fifty or more qubits have allowed us to enter a computational era where the output results are not easily simulatable on the world’s biggest supercomputers.  What we have not seen yet, however, is whether or not such quantum complexity can be ever useful for any practical applications. 
A fundamental question behind this lies in the non-trivial relation between the complexity and its computational power.  If we find a clue for how and what quantum complexity could boost the computational power, we might be able to directly utilize the quantum complexity to design quantum computation even with the presence of noise and errors.  
In this work we introduce a new reservoir computational model for pattern recognition showing a quantum advantage utilizing scale-free networks. 
This new scheme allows us to utilize the complexity inherent in the scale-free networks, meaning we do not require programing nor optimization of the quantum layer even for other computational tasks.  The simplicity in our approach illustrates the computational power in quantum complexity as well as provide new applications for such processors.
\end{abstract}

\maketitle

\section{Introduction}
The recent realization of quantum processors with fifty plus qubits is undoubtedly a key milestone for the nascent quantum technology field \cite{Arute2019, Zhong2020, Gong2021}. 
The quantum advantage has been estimated through its complexity generated by these quantum processors, benchmarking it against the necessary run time for a classical computer to simulate it.  Such a complexity has been believed to be deeply connected to its quantum computational power, yet it still remains unclear in what way we could extract usable computational power for real applications from the quantum complexity itself.

Quantum neural networks (QNNs) have been considered as one of the potential directions to exploit quantum complexity that the current quantum processors (including Noisey Intermidate Scale Quantum (NISQ) devices) can generate.  There has been a number of approaches and definitions for QNNs \cite{Schuld2014,Ezhov2000}.  In the early days of QNNs, the core ideas behind perceptron (a model of artificial neural networks) had been extended to QNNs \cite{Altaisky2001}, where the perceptron activation function was replaced by an operator.  Despite the simple notion of the quantum perceptron given by Altaisky, its implementation remains highly nontrivial.  The main focus of the early works on quantum perceptron is to integrate the nonlinearity present in the classical model into quantum mechanics \cite{Guta2001, Panella2011, Sagheer2013, Zhou2007}.
More recently QNNs have been extensively investigated as a subclass of variational quantum algorithms (VQAs) \cite{Abbas2021,Noori2020}.  VQAs can be modeled with a feature map and a variational model with classical feedfoward, some of which have been demonstrated on gate-based quantum computers \cite{Havlicek2019, Xia2021}.  It has been suggested that some quantum neural networks may offer an advantage over classical neural networks \cite{Abbas2021}, however the optimization of a parameterized quantum circuit with a classical feedback loop would be difficult to scale \cite{McClean2018}.  Quantum reservoir computation (QRC) can also be considered as a type of QNNs, which are introduced to mainly analyse temporal/sequential data processing with both qubits \cite{Fujii2017, Martinez2021} and continuous variables \cite{Govia2021}.  In those works quantum neural networks are considered in the Hilbert space, whereas another approach uses a network of physical qubits \cite{Martinez2020}, which has been used to construct a universal quantum reservoir computer \cite{ghosh2020}.

In this paper, we propose a new quantum reservoir computation (QRC) model based on scale-free networks that emerge during the melting of a discrete time crystal \cite {estarellas2019simulating} and apply it to a pattern recognition task.  
These networks represent the effective Hamiltonian of a periodically driven system obtained through the application of tools in Floquet theory \cite{Bastidas2018}.  In our model, the classical {\it reservoir}, that is a large nonlinear system, is replaced by a quantum hidden layer characterized by the non-trivial complexity of scale-free networks.  Non-trivial complexity as well as randomness has played an important role in both quantum and classical neural networks \cite{Kinouchi2006, Martinez2021}.
\begin{figure*}[t]
\centering
\includegraphics[width=\textwidth]{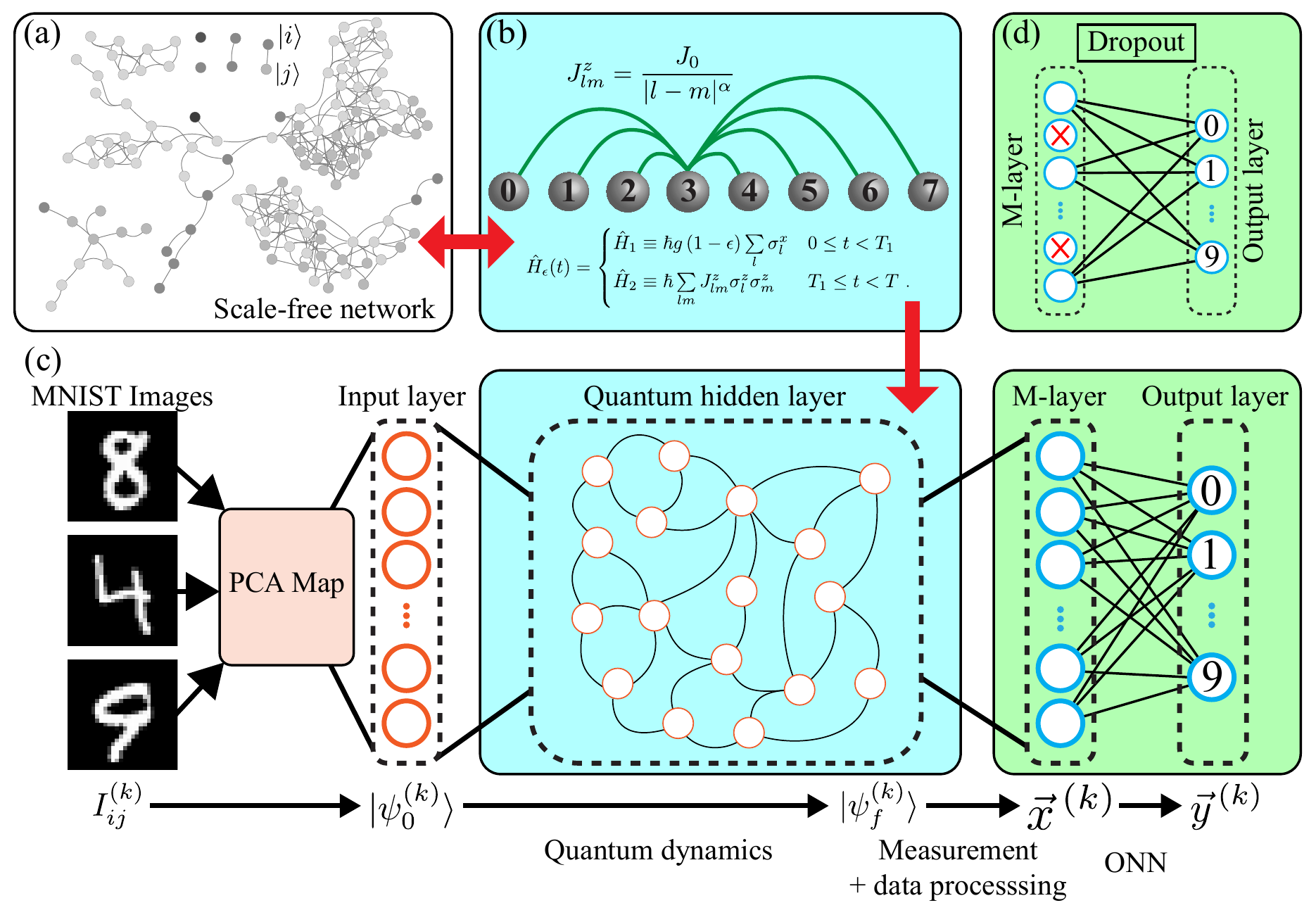}
\caption{ A schematic illustration of the structure of the quantum reservoir computer.  (a) shows an example of scale-free networks represented in the Hilbert space. Each dot indicates a computational basis state, while each edge is weighted with the hopping strength between the two states.  When the hopping strength is smaller than a threshold given by the percolation rule, there is no edge between the states.  (b) indicates a quantum system used to implement the quantum hidden layer.  Here the model of a discrete time crystal is used.  (c) illustrates the total scheme of our QRC model.  There are four layers: the input layer, the quantum hidden layer, the M-layer and the output layer.  The input layer encodes the input data on to the quantum hidden layer, while the M-layer is to measure the quantum system in the computation basis, converting the quantum data (state) to the classical data.  The one-layer neural network (ONN) is responsible for the training.}
\label{Fig1}
\end{figure*}

Our QRC model presents several advantages.  First and foremost the network grows exponentially with the number of qubits $N$; 50 qubits could generate a network as large as the neural network in a human brain.  Second, despite this rapid growth in complexity, the parameter setting for the quantum hidden layer remains simple.  In fact, there is no programing nor optimization necessary for the quantum hidden layer once it has been set, even for different computational tasks.
Third, by eliminating the costly parameter optimization over the variational model in VQAs, we can expect a high feasibility of our QRC model and a significant speed up in the training process.   
This advantage in the learning speed is similar to what we expect with the extreme learning machine in the classical machine learning \cite{Huang2006}.

\section{The model}
Our quantum reservoir computation (QRC) model consists of four layers as illustrated in Fig.\ref{Fig1}.  The computation starts with encoding the classical data $I^{(k)}_{ij}$  for the $k$-th image onto the initial state of the quantum hidden layer.  The quantum hidden layer is the {\it reservoir} of this computational model.  It may not be necessary, however it is desirable to preprocess the classical input data so that the encoding can be done easily and efficiently.  For this we chose the technique of Principle Component Analysis (PCA), as indicated in Fig.~\ref{Fig1}(c).  After some time in the dynamics of the quantum hidden layer the output quantum state $|\psi_f^{(k)}\rangle$ is measured in the computational basis at the M-layer, converting the quantum data $|\psi_f^{\,(k)}\rangle$ to the classical data $\vec{x}^{\, (k)}$.  The classical data is then renomalized for the one-layer neural network (ONN), that is a classical neural network only with the input and output layers.  Training is performed only at the output layer.

Since the quantum hidden layer is the solo quantum component in our scheme, how it is designed is key to achieve a quantum advantage.  
The quantum hidden layer (Fig.\ref{Fig1} (b)) and its network (Fig.\ref{Fig1} (a)) are connected through the visualization of the Hamiltonian \cite{Bastidas2018}.  The Hamiltonian of an $N$-qubit system in the computational basis $\{|i\rangle\}$ where $i$ is a binary digit from $0$ to $2^N-1$ may be represented as $H=\sum_i  \mathcal{E}_i |i \rangle \langle i| + \sum_{i,j (i\neq j)} \mathcal {W}_{ij}|i\rangle \langle j|$ where $ \mathcal{E}_i$ and $\mathcal {W}_{ij}$ correspond to the energy of the computational basis $|i\rangle$ and the transition energy between $|i\rangle$ and $|j\rangle$ respectively.  The network has an edge between the $i$-th node (state) and $j$-th node (state) with weight $\mathcal{W}_{ij}$ when the percolation condition $| \mathcal{E}_j-\mathcal{E}_i|<| \mathcal{W}_{ij}|$ is satisfied \cite{Bastidas2018}.  The percolation rule is to eliminate the off-resonant transitions from the network. 
In the case of the periodically driven system, the network is generated from an effective Hamiltonian using the Floquet theory.  

There are now two approaches to setting the quantum hidden layer in the targeted regime.  Using a network with the appropriate complexity (Fig.\ref{Fig1} (a)) as a recipe, we can generate a Hamiltonian with the scale-free nature by assigning each computational basis state to each node of the network.  The Hamiltonian $H$ can be implemented as the time evolution $e^{-iH\tau/ \hbar}$ of the system.  It is however likely that such a Hamiltonian involves many-body interactions or the gate decomposition of $e^{-iH\tau/ \hbar}$ can be computationally costly.  Another approach is to find a quantum system with such a Hamiltonian, and the melting of a discrete time crystal is a known example of such quantum systems.\cite{estarellas2019simulating}.

The Hamiltonian of our discrete time crystal $H_{\epsilon}(t)$ is given by \cite{Zhang2017,Yao2017}
 \begin{equation}
         \label{hamiltonian}
  \hat{H}_{\epsilon}(t)=
  \begin{cases}
   \hat{H}_1 \equiv \hbar g \left(1 - \epsilon \right)\sum_{l} \sigma_{l}^{x} & 0 \leq t < T_1 \\
   \hat{H}_2 \equiv \hbar\sum_{lm}J_{lm}^{z} \sigma_{l}^{z} \sigma_{m}^{z} & T_1 \leq t < T \ .
  \end{cases}
   \end{equation}
Here $\{\sigma_{l}^{x},\sigma_{l}^{y},\sigma_{l}^{z}\}$ are the Pauli operators on the $l$-th qubit, while $J_{lm}^{z} \equiv J_0/|l-m|^{\alpha}$ represents the long-range interaction between the $l$-th and $m$-th qubits that takes the form of an approximate power-law decay with a constant exponent $\alpha$. Next the parameter $g$ satisfies the condition $2gT_1=\pi$ such that $\hat{U}_1=\exp{\left( -\mathrm{i}\hat{H}_1 T_1/\hbar\right)}$ becomes a global $\pi$ pulse around the $x$-axis.  We assume $T_1=T/2$ for convenience.   The parameter $\epsilon$ is our rotation error \cite{estarellas2019simulating}.
Because this is a periodic system, the Floquet operator is given by
\begin{equation}
         \label{eq:FloquetOperatorTimeCrystal}
         \hat{\mathcal{F}}_{\epsilon}=\hat{U}_{\epsilon}(T)=\exp{\left( -\frac{\mathrm{i}}{\hbar}\hat{H}_2 T_2\right)}\exp{\left( -\frac{\mathrm{i}}{\hbar}\hat{H}_1 T_1\right)}
         \ .
\end{equation}
The network representation can be obtained from the effective Hamiltonian of the Floquet operator $\hat{H}^{\text{eff}}_{\epsilon}=i \hbar/T \log[\hat{\mathcal{F}}_{\epsilon}]$ and the percolation rule.  When $\epsilon=0$, a Period-$2$ Discrete Time Crystal (2T-DTC) appears, and the corresponding network is a locally-connected network (a set of dimers).  
As $\epsilon$ gets larger, the network starts to grow following a preferential attachment mechanism and typically forms a scale-free network, and when  $\epsilon$ reaches a critical region, the network goes through the transition from scale-free to random \cite{estarellas2019simulating}.  In our QRC, we intend to set $\epsilon$ to $0.03$ for a near optimal computation, which corresponds to the transition regime.

The discrete time crystal has been experimentally demonstrated \cite{Choi2017, Zhang2017, Frey2021}, and the computational time we set for the quantum hidden layer is shorter than half of the typical coherence time of an ion trap experiment with 10 qubits \cite{Zhang2017}.   The discrete time crystal can also be implemented on a gate-based quantum processor by a gate decomposition of the Floquet operator or the entire unitary map for the quantum dynamics \cite{Frey2021}.  Hence the model can be adapted to various quantum computational systems.

\section{Pattern Recognition with QRC}
\begin{figure}[htb]
\centering
\includegraphics[width=0.45\textwidth]{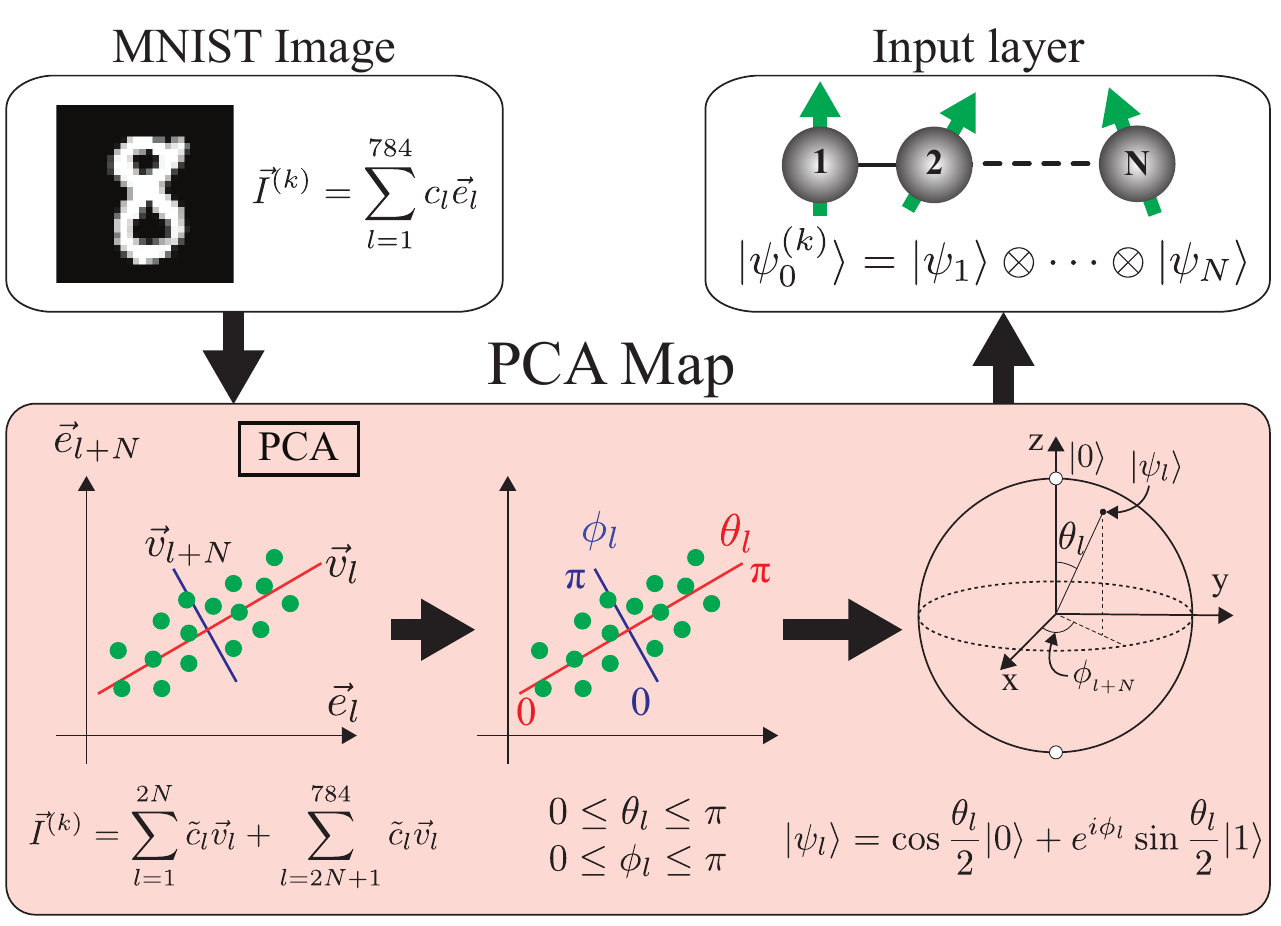}
\caption{ Schematic diagram detailing how the $2N$ parameters are extracted from a sample image and encoded onto $N$ qubits.  The left top box shows that the $k$-th image from the MNIST Image set requires 784 parameters to represent the data in full.  In the PCA Map (the bottom box), the PCA is used to select the $2N$ most influential parameters. and a pair of parameters $(c_l, c_{\tiny{l+N}})$ is encoded on one qubit by single qubit rotations.  Panel (c) shows the input state of the quantum hidden layer.
}
\label{Fig2}
\end{figure}

Here we illustrate how the QRC model performs pattern recognition.  We use the MNIST handwritten digit data set publicly available \cite{MNIST} to benchmark the performance.  The data set contains 70000 ($28\times28$ pixel) images of handwritten digits between 0 and 9.  The classical data for each image is first processed by the Principle Component Analysis (PCA), as shown in Fig.\ref{Fig2}, which allows us to select elements from the largest contribution.  With the PCA, the $k$-th image data is represented as a vector $\vec{I}^{\,(k)} =  \sum_{j=1}^{784} \tilde{c}_j \vec{v}_j$, where the $j$-th element is the $j$-th contribution among the $784$ elements.  Although it is possible to encode all $784$ values to the quantum input state, it may require a complicated quantum circuit, instead we select the $2N$ largest contribution elements to encode to the quantum input state.  These $2N$ values are encoded to each qubit by single-qubit rotations only; for the $l$-th qubit, the mapping is $\tilde{c}_l \to \theta_l$ and $ \tilde{c}_{N+l} \to \phi_l $ and the quantum input state is $|\psi_l\rangle= \cos \frac{\theta_l}{2}|0\rangle + e^{i\phi_l}\sin\frac{\theta_l}{2}|1\rangle$. 

Our previous work \cite{estarellas2019simulating} has identified the value of $\epsilon$ that marks the transition regime from scale-free to random networks, and we choose $\epsilon$ from  that parameter regime.  For our numerical simulations we set our parameters to be $J_0T=0.06$, $\alpha=1.51$.  The coupling strength corresponds to the weak coupling regime.

Now in the M-layer, we measure the output state $|\psi_f^{\,(k)}\rangle$ in the computational basis and extract the classical data.  As each measurement does not provide the entire information of the quantum state, we build up the distribution at the M-layer by repeating the process.  We note that there is no feedforward in these repeating processes.  The distribution is then renomalized to have an average $0$ and variance $1$ for the convenience of the classical data analysis.  The renomalized distribution is the output of the M-layer $\vec{x}^{\,(k)}$.

The output $\vec{x}^{\,(k)}$ is then weighted with the weight matrix $W$ of the ONN.   We follow a method widely used in pattern recognition (see appendices for details) to obtain the output of the computation $\vec{y}^{\,(k)}$.

\section{Performance and Properties}
\begin{figure}[htb]
\centering
\includegraphics[width=0.45\textwidth]{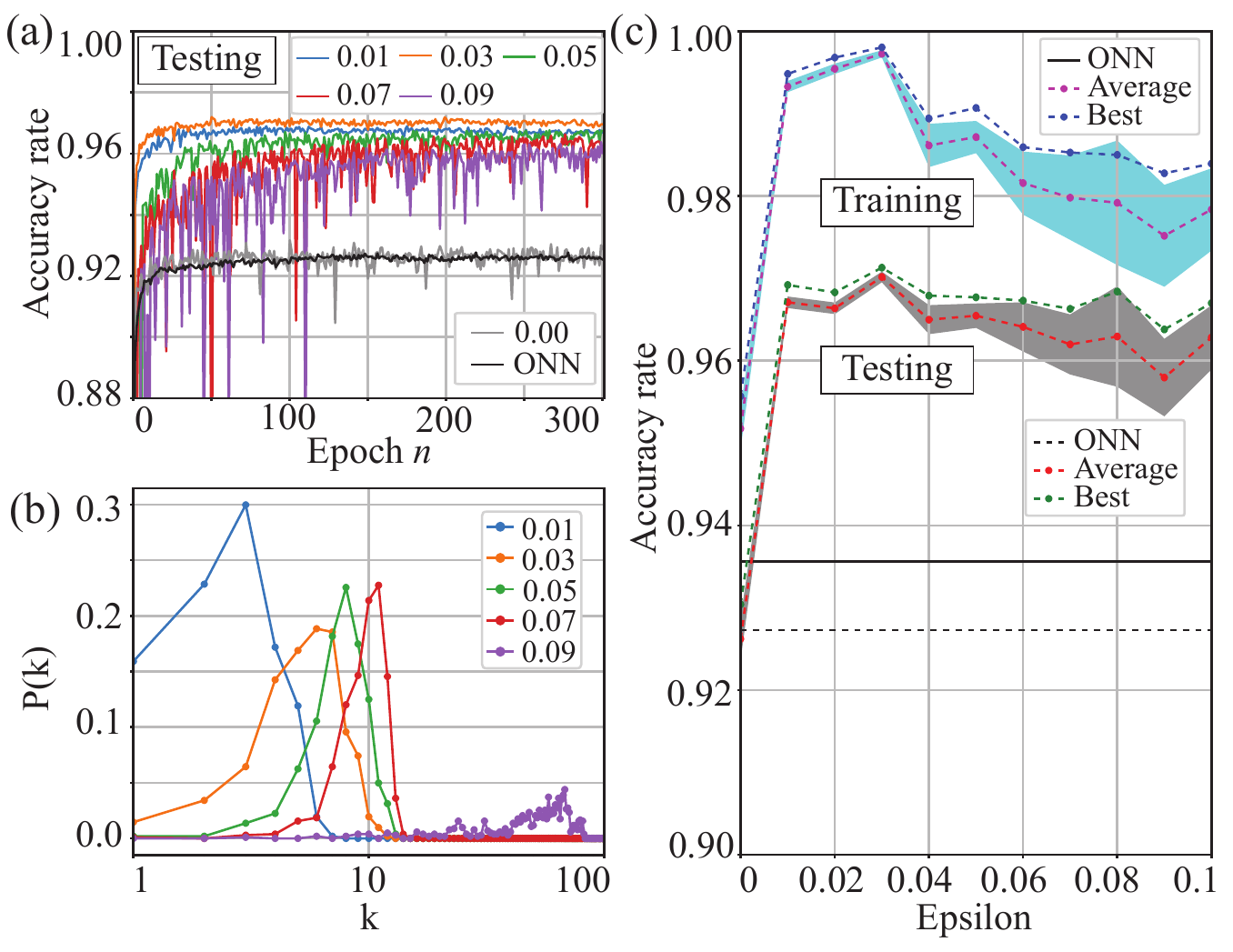}
\caption{Performance of the QRC.  Plot (a) shows the accuracy rate versus the number of epoch for each value of $\epsilon$.  The number of periods for the time evolution of the quantum hidden layer $n$ is $50$, and the coupling strength $J_0T=0.06$.  The degree distribution of the network for the efficient Hamiltonian for each $\epsilon$ is plotted in (b), where $k$ is the degree (the number of links) at each node and the P(k) is the number of node for the degree of $k$.  The plot (c) shows the average and best accuracy rates with its standard deviation for both training and testing respectively.  The average and the standard deviation are taken for 200 to 300 epochs.  The number of qubits is $N=11$.
}
\label{Fig3}
\end{figure}

We begin by first training our ONN with 60,000 samples using gradient decent, back propagation and the mini-batch method, where we evaluate the accuracy rate for the test samples.  To evaluate the optimal value of $\epsilon$, we calculate the $\epsilon$-dependency of the performance for both training and testing.  Fig.~\ref{Fig3} shows the accuracy rate for testing against various values of $\epsilon$.  The system size of the quantum hidden layer is $N=11$.

In Fig.\ref{Fig3}, we compare the performance of our QRC model for various values of $\epsilon$.  The black line in Fig.\ref{Fig3} (a) shows the accuracy rate for the QRC model without the quantum hidden layer (that is only the ONN) with the full classical input data ($784$ PCA elements).   The gray line is for $\epsilon=0$, and in this case we may not necessarily expect a quantum advantage from the quantum dynamics, however it is interesting to observe that the reduction of the classical information at the input layer does not significantly affect the accuracy rate, which indicates that the mapping between the classical data and the quantum state as well as the application of PCA helps to push up the accuracy rate.  
In Fig.\ref{Fig3} (a), the accuracy rate for testing quickly goes up high from the $\epsilon$ value from $0$ to $0.03$.  The optimality peaks around $\epsilon=0.03$ and gradually reduces when $\epsilon$ gets larger.  The degree distributions of the network for the effective Hamiltonian with $\epsilon=0.01$ to $0.03$ exhibit the scale-free nature as shown in Fig.\ref{Fig3} (b).  The non-trivial complexity in the quantum hidden layer contributes to the computational power in this model for both the speed of the leaning and the accuracy rate.  This comparison with the network property and the accuracy rate suggests that we need to design complex behavior of the quantum hidden layer to extract the computational power for the task at hand and the network analysis could be a useful tool to design the quantum hidden layer. 
Fig.\ref{Fig3} (c) shows the $\epsilon$-dependency on the accuracy rate for training and testing respectively.  

\begin{figure}[htb]
\centering
\includegraphics[width=0.45\textwidth]{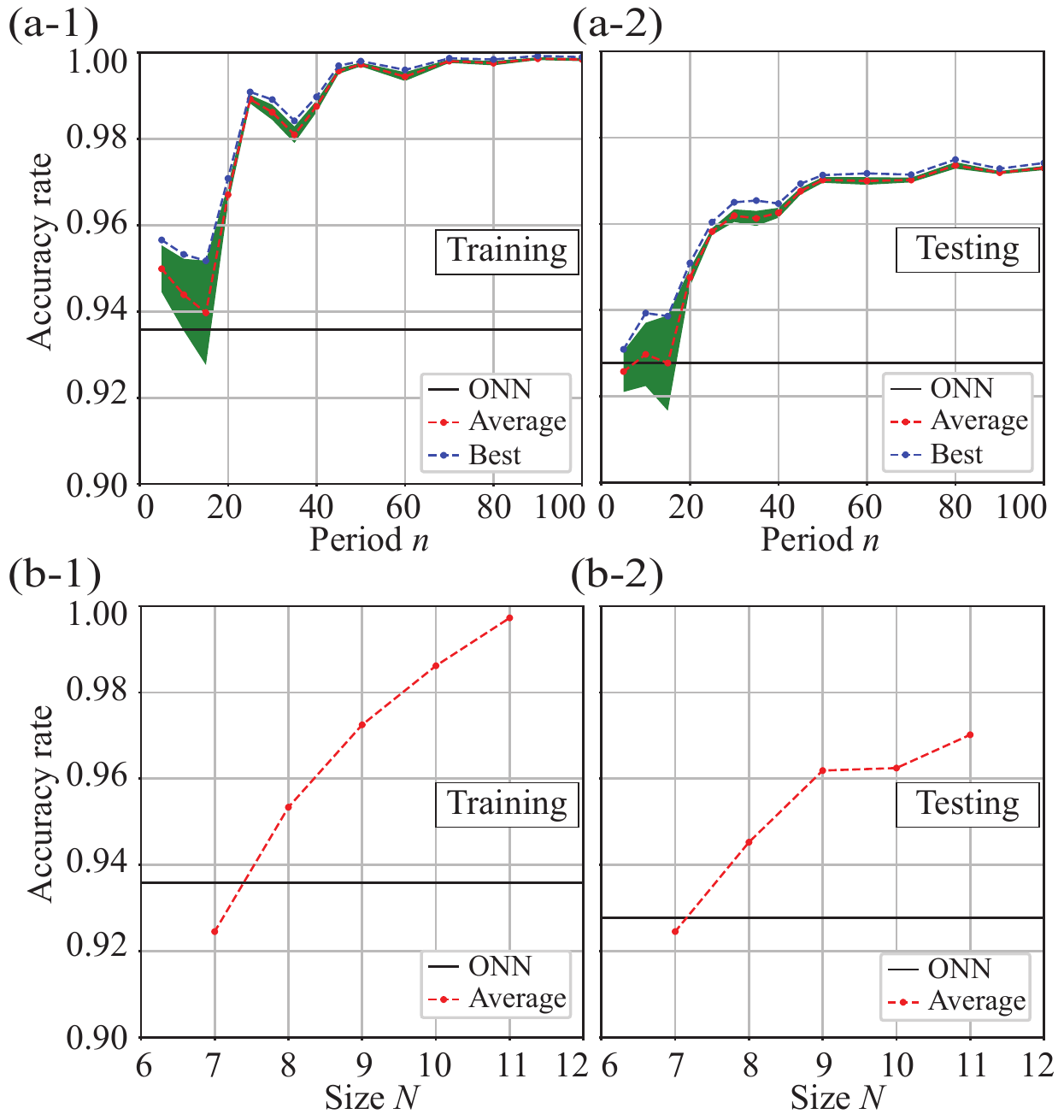}
\caption{ Performance of the QRC with time ($nT$) and size ($N$).  The top two figures plot the average and best accuracy rates with its standard deviation for training (a-1) and testing (a-2) for the different periods with $N=11$.  The bottom two figures show the size dependency of the accuracy rate for training (b-1) and testing (b-2).  The parameters are set as $J_0T=0.06$ and $\epsilon=0.03$.  The average and the standard deviation are taken for 200 to 300 epochs. 
}
\label{Fig4}
\end{figure}

In the above evaluation, the number of periods $n$ is set to $50$ meaning the time duration $nT$.  If $n$ is too small, the quantum hidden layer would not be able to exploit its large Hilbert space.  Figs.~\ref{Fig4} (a) show the $n$-dependency of the accuracy rate for both training (a-1) and testing (a-2). The accuracy rates almost saturate around $n=50$, which is feasible based on current ion-trap technology \cite{Zhang2017}. 

Next, we analyze the size effect of the quantum hidden layer on the accuracy rates.  When increasing $N$, we slightly increase the dimension of the Hilbert space of the quantum hidden layer as well as the input of the PCA elements.  Figs.~\ref{Fig4} (b) show that the accuracy rate for training and testing.  In both cases the accuracy rate goes up with a larger system size $N$.  In fact, only two extra qubits  from $N=7$ to $9$ gives a nearly $4\%$ improvement in the accuracy for testing.   

The scale free nature of the Hamiltonian previously investigated is statistical with an ensemble of disorder distribution on qubits, and hence each instance could be deviated from the typical power-law degree distribution (a feature of scale-free networks).  In our system, when $N$ is odd, the degree distribution shows the power-law nature, though for even $N$s it tends to be less typical.  This difference arises from the fact that the system with the flat disorder distribution holds additional system symmetries when $N$ is even.  Fortunately, our numerical results show that the performance of our QRC model is not sensitive to the actual degree distribution as long as $\epsilon$ stays in the regime, which eliminate the necessity to check the degree distribution of each network.

\section{Overfitting and Dropout}
\begin{figure}[htb]
\centering
\includegraphics[width=0.40\textwidth]{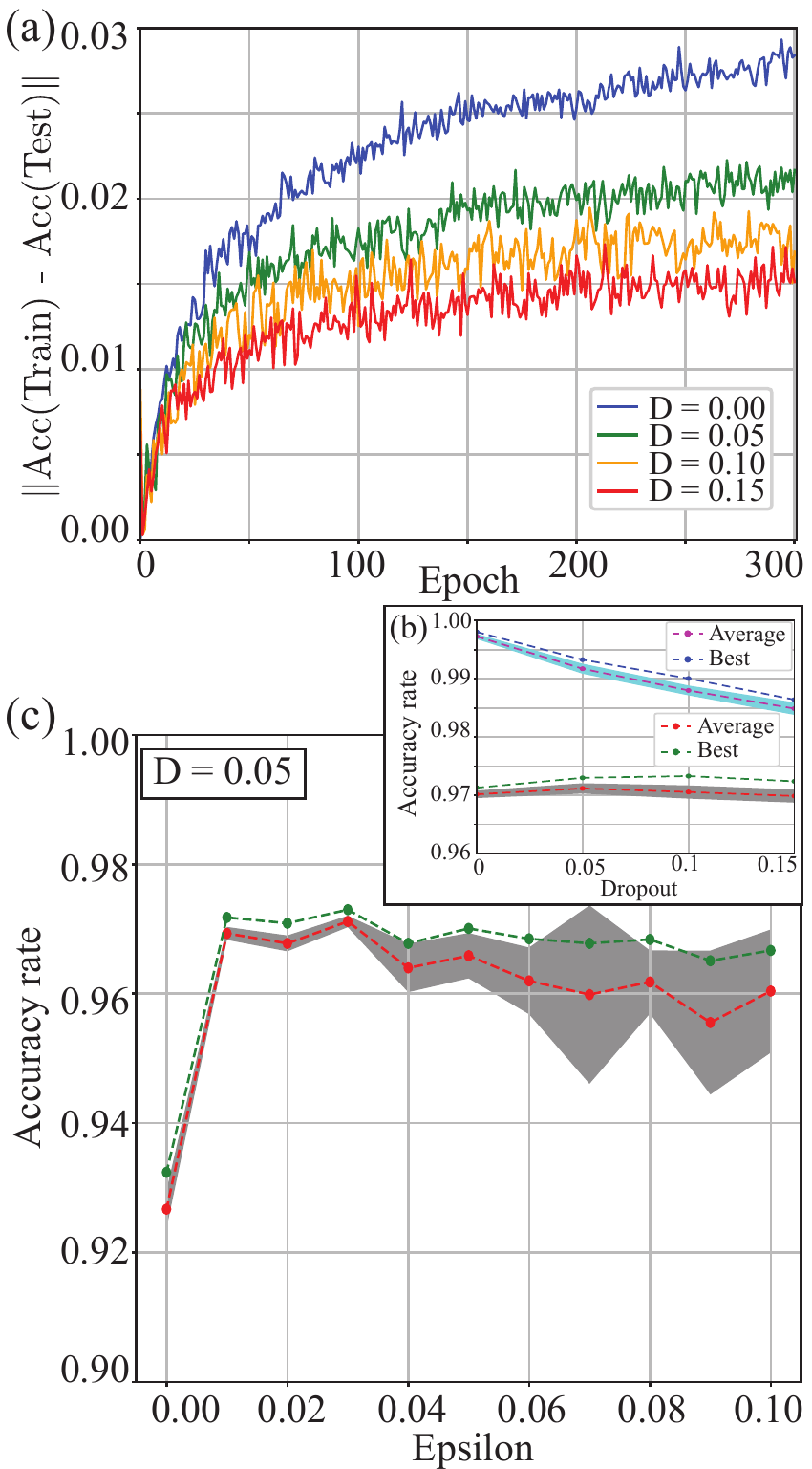}
\caption{ The effect of drop out on the accuracy of the QRC.  Figure (a) shows the accuracy rate difference between training and testing for each dropout.  The difference for all nonzero dropout rates saturates as the learning proceeds, which indicates that overfitting can be suppressed by dropout. Figure (b) shows the accuracy rates for training (top lines) and testing (bottom lines) to choose the best rate for the dropout, which is $D=0.05$ $(5\%$) in this case.  Overfitting can be further suppressed by increasing $D$, however the overall performance starts to drop for a large $D$.  Figure (c) shows the $\epsilon$-dependency on the accuracy rate for testing with $5\%$ dropout.  All other parameters are the same as in Fig.~\ref{Fig3}.
}
\label{Fig5}
\end{figure}

One of the most common problems with machine learning is overfitting, and our QRC model is no exception.  The gap between the accuracy rates for the training and testing samples in Figs.~\ref{Fig4} (a) gets larger as the learning progresses, and the accuracy rate for training in Fig.\ref{Fig4} (b-1) almost reaches unity at $N=11$, which suggests overfitting. 
To capture this more clearly, we plot the difference between the accuracy rates for training and testing in Fig.\ref{Fig5} (a).  Here we observe that the blue line does not saturate.  To address this, we introduce Dropout to the ONN in our model as shown in Fig.~\ref{Fig1}(d).  The dropout is one of the techniques developed in classical machine learning to circumvent overfitting by randomly erasing information in neural networks.  It is not commonly used due to the computational cost associated with the technique \cite{Srivastava2014}, however in our case, the cost of dropout is minimal with the one-layer neural network. 
We plot the effect of dropout in our QRC model in Fig.\ref{Fig5} (a), by showing the difference between the accuracy rates for four different degrees of dropout ranging from $0 - 15\%$.  When the dropout is introduced, the overfitting is significantly suppressed.  

Now we re-evaluate the accuracy rates for the QRC model, which are plotted in Fig.~\ref{Fig5} (b) and (c).  Fig.\ref{Fig5} (b) shows that we can achieve the best performance for testing with $5\%$ dropout in this case, while the Fig.~\ref{Fig5} (c) shows the accuracy rate for testing samples with $5\%$ Dropout for different values of $\epsilon$. The performance of the QRC model slightly increases with the dropout as shown in (b), and hence dropout successfully suppresses the accuracy rate for training and increases it for testing for a larger $N$ where the overfitting is more severe.

\section{Conclusion}
We have shown the potential for quantum advantage in the QRC model for pattern recognition.  Our QRC model is the first example to directly facilitate quantum complexity for computation without programing, nor quantum/classical feedforward.  The quantum hidden layer can be considered as an accelerator or booster to be inserted into classical neural networks, and for the pattern recognition task in this paper, it has been shown that a quantum hidden layer as small as 8 qubits can successfully boost the computational power.  As the dynamics of the quantum hidden layer is fixed, no re-programing is required for the quantum hidden layer to adopt different tasks, and all we need to do is to upload the new parameter setting for the ONN and the input-state preparation protocol.  Similarly, the quantum hidden layer can be inserted into quantum neural network, in particular it might be a good candidate for feature map in VQA, as recently observed that the complexity of feature map has a crucial role for VQAs to show a quantum advantage~\cite{Abbas2021}.
 
The technology requirement to implement our QRC model is modest, and feasible based on the current technology.  There are several different implementations possible for the effective Hamiltonian we used in this paper: the periodically driven system is one example, and gate-model quantum computation is another.  As the key to achieve the acceleration lies in complexity of quantum dynamics, we may also employ different quantum models.  Quantum cellar automata could be a simple extension of our case, and boson sampling, though not universal, can be a feasible candidate.   The simplicity of our scheme has paved a way to a new platform of practical quantum computation.

\section{Acknowledgements}
We thank  Victor M.~Bastidas and Aoi Hayashi for valuable discussions. 
This work was supported by MEXT Quantum Leap Flagship Program (MEXT Q-LEAP) Grant Number JPMXS0118069605.

\appendix
\section{Principle Component Analysis}
In our model, Principle Component Analysis (PCA) is used to effectively encode the classical information to the quantum hidden layer as shown in Fig.\ref{Fig2}. The $k$-th MNIST image is converted to the vector $\vec{I}^{\, (k)}$.  The basis vectors $\{\vec{v}_j\}_{j=1,\ldots, 784}$ for this vector representation need to be optimized with the sample data.  Once the basis vector set has been optimized with the sample data, the same set is used to represent the test images.
The conversion $\tilde{c}_l \to \theta_l $ or $ \phi_l $ is given by,
\begin{equation}
	\theta_{l} \, \text{or} \, \phi_{l} =  \frac{ \pi \left( \tilde{c}_l - \min{ \left[\tilde{c}^{(\text{train)}}_l \right]} \right) }{\max{\left[\tilde{c}^{(\text{train)}}_l \right]} - \min{\left[\tilde{c}^{(\text{train)}}_l \right]}},
\end{equation}
where $\max{[\tilde{c}^{(\text{train)}}_l ]}$ and $\min{[\tilde{c}^{(\text{train)}}_l ]}$ mean the maximum and minimum values of $\tilde{c}_l$ across all the training samples respectively. Here, we note that for the case where $\theta_{l}$ or $\phi_{l}$ goes beyond the range $[0,\pi]$ with testing samples, we truncate the value. 
Due to the optimization of the basis vectors, the loss of information by the reduction from the $784$ elements to the $2N$ highest contribution elements is minimized.
\newline

\section{ONN with the M-layer and the output layer}
The ONN is a one-layer neural network of the M-layer with $m$ active neurons and the output layer with $n$ active neurons and is to be optimized.  

The number of neurons at the M-layer can in principle be as large as $2^N$, however when we apply the dropout we randomly eliminate the active neurons in this layer as illustrated in Fig.\ref{Fig1} (d).  At the output layer, there are ten neurons, each corresponding to a digit to be recognized. 

At the output layer the effects of the weight matrix is summarized and a shift $\vec{B}$ is applied as
\begin{equation}
	 u_l=\sum_{i=1}^n x_i\cdot w_{i,l}+B_l.
 \label{eq:outputlayer}
 \end{equation}
where $x_i$ is the $i$-th element of the output of the M-layer $\vec{x}$ and and $b_l$ is the $l$-th element of the shift $\vec{B}$.
As our computational task is pattern recognition, we insert the activation function $f$ to convert the data $\vec{u} $ to $\vec{y}$ as $\vec{y}= f(\vec{u})$.  

For the activation function we employ the soft-max function~\cite{Haykin2010,Goodfellow2016}, which is widely used in classification problems~\cite{Haykin2010,Goodfellow2016}, 
\begin{equation}
    y_l = f(x_l) =  \frac{\exp(x_l)}{\sum\limits_k \exp(x_k)}.
    \label{eq:acctivef}
\end{equation}
We define the lost function $L_k$ for the $k$-th sample by the cross entropy as
\begin{equation}
    L_{k} = - \sum_l^{n} t^{({k})}_l \log(y^{({k})}_l),
    \label{eq:cosf}
\end{equation}
to evaluate the learning progress.  Here $\vec{y}^{\,({k})}$ is the output vector, while $\vec{t}^{\,\, (k)}$ is a one-hot vector, that is a basis unit vector of the ten dimensional vector space. The one-hot vectors represents the correct result and hence are the reference for the evaluation of the output of the ONN.
Then we apply Gradient Descent, Back Propagation, and the Mini-batch Method to optimize the ONN.

\section{Gradient Descent and Back propagation }
The weight matrix ($w_{ij}$) and the shift vector ($B_{i}$) are optimized using the Gradient descent method, through 
\begin{equation}
\begin{split}
 w_{ij}^{(n+1)} &= w_{ij}^{(n)} - \eta  \frac{\partial L_l }{\partial w_{ij} } \\
 b_{i}^{(n+1)} &= b_{i}^{(n)} - \eta  \frac{\partial L_l }{\partial b_{i}},
\end{split}
\end{equation}
where $\eta$ is a learning rate. To calculate the derivatives, we use the back propagation method. 
Applying the chain rule with Eqs.~\cref{eq:outputlayer,eq:acctivef,eq:cosf}, we have
\begin{equation}
    \begin{split}
        \frac{\partial L_l }{\partial w_{ij} } &= x_{l,i}(y_{j}^{(l)}-t_{j}^{(l)}) \\
        \frac{\partial L_l }{\partial b_{i}} &= (y_{j}^{(l)} - t_{j}^{(l)})
    \end{split}
\end{equation}

\section{Mini-batch Method}
We use the mini-batch method to reduce the computational cost and to avoid local minima.
In this method, the loss function is now the average of the loss function for each sample and is given by
\begin{equation}
    L = \frac{1}{M}\sum_{l=1}^{M} L_l,
    \label{eq:mean-cost}
\end{equation}
where $M$ is the batch size and the $L_l$ is the cost function for the $l$-th sample.  The hidden layer can be written as the matrix, 
\begin{equation}
    X = \left ( 
    \begin{matrix}
        \vec{x_1}^T\\
        \vdots \\
        \vec{x_M}^T
    \end{matrix}
    \right).
\end{equation}
    
In this method, the derivative for the weight matrix is given in the matrix representation by
\begin{equation}
    \frac{\partial L }{\partial w_{ij} } = X^T\cdot 
\left(
\begin{matrix}
    \vec{y}^{\,(1)}  - \vec{t}^{\,\,(1)} \\
    \vdots \\
    \vec{y}^{\,(M)} - \vec{t}^{\,\,(M)}  \\
\end{matrix}
\right).
\end{equation}
The derivative for the bias vector is then a sum of each derivatives as
\begin{equation}
    \frac{\partial L }{\partial b_i} = \sum_{l}^M \frac{\partial L_l }{\partial b_{i}}.
\end{equation}




\end{document}